# Self-Assembled Photochromic Molecular Dipoles for High Performance Polymer Thin-Film Transistors

*Satyaprasad P. Senanayak[1,5], Vinod K. Sangwan[2], Julian J. McMorrow[2], Ken Everaerts[3], Zhihua Chen[4], Antonio Facchetti[2,4]\*, Mark C. Hersam[2,3]\*, Tobin J. Marks[3]\* and K.S.Narayan[1]\**

Affiliation:

[1]Chemistry and Physics of Materials Unit
Jawaharlal Nehru Centre for Advanced Scientific Research
Bangalore 560064, India

[2]Department of Materials Science and Engineering, and the Materials Research Center
Northwestern University
Evanston, Illinois 60208, United States

[3]Department of Chemistry, and the Materials Research Center
Northwestern University
Evanston, Illinois 60208, United States

[4]Flexterra Inc.
8025 Lamon Avenue
Skokie, Illinois 60077, United States

[5]Optoelectronics Group
Cavendish Laboratory, University of Cambridge,
JJ Thomson Avenue, Cambridge, CB3 0HE, UK

**Abstract:**

The development of high-performance multifunctional polymer-based electronic circuits is a major step towards future flexible electronics. Here, we demonstrate a tunable approach to fabricate such devices based on rationally designed dielectric super-lattice structures with photochromic azo-benzene molecules. These nanodielectrics possessing ionic, molecular, and atomic polarization are utilized in polymer thin-film transistors (TFTs) to realize high performance electronics with p-type field-effect mobility ($\mu_{FET}$) exceeding 2 cm$^2$V$^{-1}$s$^{-1}$. A crossover in the transport mechanism from electrostatic dipolar disorder to ionic-induced disorder is observed in the transistor characteristics over a range of temperatures. The facile supramolecular design allows the possibility to optically control the extent of molecular and ionic polarization in the ultra-thin nanodielectric. Thus, we demonstrate a three-fold increase in the capacitance from 0.1 µF/cm$^2$ to 0.34 µF/cm$^2$, which results in a 200% increase in TFT channel current.

**Introduction**

Polymer field effect transistors (PFETs) are of great interest in the development of low-cost, flexible electronics for sensors, smart cards, and non-volatile memories.[1] Ongoing efforts in the design of molecular dielectrics[2] and semiconductors,[3-5] innovation in device geometries,[6-7] as well as thin-film layer fabrication procedures[8] and growth conditions[9-10] have resulted in enhanced PFET performance with field-effect mobility ($\mu_{FET}$) > 1 cm$^2$V$^{-1}$s$^{-1}$, switching speed > 1 MHz, and I$_{on}$/I$_{off}$ ~ 10$^6$.[6, 11-14] Furthermore, polymer electronics provides an avenue for embedding multiple functionalities into a device structure for photo-detectors,[15] light emission,[16-17] chemical sensors,[18] and memory applications.[19] The general strategy for introducing optical functionality in PFETs is to either directly photoexcite the semiconducting polymer layer or to excite the dielectric layer blended with photochromic materials.[20-22] Alternatively, interesting techniques such as the introduction of self-assembled photochromic (SAP) layers at the electrode or semiconductor-dielectric interface[23-24] have also been utilized to modulate device properties such as interface disorder and contact resistance.[25-29] However, in the majority of cases, the self-assembled layers are restricted to a single monolayer.[30-31] As a result, light-modulated photochromic processes can modulate the channel current ($\Delta I_{ds}$) by less 20% in most reported systems.[21] Towards this end, here we demonstrate the use of solvent-stable self-assembled nanodielectrics (SANDs) based on alternating organic and inorganic layers that exhibit polarization variation with optical and electrical stimuli.[32] The photochromic properties of the organic layer[33-35] in SAND with a PFET geometry are used to tailor the dielectric capacitance and interface trap density in the device to reversibly tune the drain current $\Delta I_{ds}$ by up to 200 %.

In a PFET, the charge transport occurs in the semiconducting channel at the semiconductor-dielectric interface and thus, the polarization of the dielectric layer significantly modifies the transistor characteristics. In general, high-$k$ inorganic dielectrics such as $Al_2O_3$, $HfO_2$, $Ta_2O_5$ or organic dielectrics such as PVDF based on molecular dipoles, modify the charge transport in PFETs unfavorably due to the development of a polarization cloud that increases the effective mass of the charge carriers.[36-41] In contrast, dielectrics based on ionic polarization, such as electrolytes or ionic liquids; have low interfacial disorder due to increased charge density and charge screening, resulting in enhanced field-effect mobilities.[42-43] In this study, we utilize hafnia-based SAND dielectric structures consisting of inorganic $HfO_x$ layers and organic π-conjugated based 4-[[4-[bis(2-hydroxyethyl)amino]-phenyl]diazenyl]-1-[4-(diethoxyphosphoryl) benzyl]pyridinium bromide (PAE) layers having $Br^-$ ions. The overall polarization in the SAND structures has contributions from the lattice dipoles of $HfO_x$, molecular dipoles of the PAE groups, and ionic polarization from the bromide anions present in the superlattice.[2, 44-45] In addition, the transport characteristics also indicate a transition from a conventional electrostatic dipolar disorder to ionic-induced disorder, as extracted from the direction of hysteresis that is tracked in the temperature-dependent transfer characteristics. We then utilize the photochromic nature of the polar PAE molecule to optically switch between the isomeric cis/trans isomers of the stilbazolium cation and analyze the effect of this isomerization on the overall dielectric constant of the SAND superlattice. This optical switching of the dielectric constant is utilized in a PFET geometry to modulate charge transport, understand the role of ionic polarization in the SAND structure, and realize photoresponsivity (R) in SAND-based PFETs as high as 2 A/W.

**Results and Discussion:**

Bottom-gate/top-contact thin-film transistor (TFT) devices were fabricated on glass substrates having Al as the gate contact (30 nm thick) coated with a SAND dielectric multilayer (see the Supporting Information for details). Both n-type (N2200) and p-type (PBTOR) semiconducting layers were next deposited by spin-coating, and the devices were completed by thermal evaporation through a shadow mask of 30 nm thick Al and Au source-drain contacts (W/L = 1000/60), respectively. Both n-FETs and p-FETs exhibit well-defined linear and saturation regime transistor characteristics at low operating voltages of $V_d$, $V_g$ = |5 V|, as shown in **Figures 1** and **S1**. Considering the fact that the transfer characteristics exhibit hysteretic behavior which is dependent on the $V_g$ sweep rate (**Figure 2**), the $\mu_{FET}$ values are estimated from the forward sweeps in the transfer curves measured at the fastest sweep rate where the hysteretic behavior is negligible. The performance parameters extracted in the saturation regime are as follows: $\mu_{FET}^p$ ~ (2.3 ± 0.3) cm$^2$V$^{-1}$s$^{-1}$, $\mu_{FET}^n$ ~ (0.1± 0.08) cm$^2$V$^{-1}$s$^{-1}$, and $I_{on}/I_{off}$ ~ $10^4 - 10^5$. It is instructive to examine the hysteresis in the $I_d(V_g,T)$ characteristics for the SAND-based PFETs to understand the transport mechanism and factors affecting the polarization of the SAND dielectric layer. Hysteresis in the transconductance characteristics of PFETs have been attributed to factors such as trapping at the interface states, arrangement of dipoles in ferroelectric systems, and/or existence of mobile ions.[46-47] This hysteretic PFET behavior can be characterized by the threshold voltage shift ($\Delta V_{th}$) and the direction of the hysteresis loop. We define the threshold voltage shift as: $\Delta V_{th} \approx V_{th,R} - V_{th,F}$ where, $V_{th,R}$ and $V_{th,F}$ are the threshold voltage in the reverse and forward sweep, respectively. Our measurements indicate a clear distinction between the observed temperature-dependent hysteretic behavior of the n-type and p-type FETs. As shown in **Figure 1**, upon increasing the temperature from 100 K to 360 K, it is observed that the area enclosed between the hysteresis curves (**Figure 1d**) and the corresponding $\Delta V_{th}$ (**Figure 1e**) increase for the PBTOR-based *p*-FETs. The clockwise direction of the hysteresis loop close to room temperature and an increase in the hysteresis with temperature is in contradiction to the typical hysteresis originating from electronic trapping.[48-52] Nevertheless, this trend is similar to the hysteresis behavior observed for PFETs having electrolyte or ionic liquid based dielectric layers[52], where the hysteresis originates from increased diffusion of ions at the interface that affects the electronic charge density at the interface.

In addition, the existence of an ionic diffusion dominated hysteresis mechanism is in the present case is also indicated from the exponential variation of the hysteresis with temperature (similar to ionic diffusion with temperature), which exhibits an Arrhenius activation energy of ~ 150 meV (**Figure 1**). The onset of the exponential variation can then be correlated with the temperature at which the ionic diffusion becomes dominant, as shown by the linear extrapolation in **Figure 1d**. Under similar conditions, the FETs fabricated with N2200-SAND exhibit no variation in hysteresis with change in temperature (**Figure S2**). Based on these observations, we conclude that the mobile ions that contribute to hysteresis are more significantly affected by negative $V_g$ than positive $V_g$. In other words, the hysteresis behavior for p-FETs originates from thermally-activated diffusion of the $Br^-$ ions, thereby increasing the observed hysteretic behavior with increasing temperature. However, in the case of n-FETs the effect of ionic hysteresis is not observed because the $PAE^+$ ions are immobilized by the inorganic $HfO_x$ layers via sigma P-O bonds, thus preventing ionic diffusion.[53-55] Considering the fact that there is no change in the off-current of the SAND-based PFETs (**Figure 1c**), note that although significant ionic polarization is observed in the hysteresis, the existence of the top $HfO_x$ capping layer prevents unwanted doping of the semiconductor. Furthermore, the presence of mobile ions at the semiconductor-dielectric interface can result in two sources of disorder: one originating from the electrostatic disorder due to traps in the dielectric, and another due to the ionic polarization at the semiconductor-dielectric interface.[56] The direction of the hysteresis sweeps can be used to track these different sources of interfacial disorder.[43, 46] In general, hysteresis originating in electrolyte or ionic liquid based PFETs has lower channel current in the forward sweep that increases in the reverse sweep mainly due to the increased density of the ions at the interface with $V_g$ bias.[42, 47] Analysis of the hysteresis direction of the present p-FETs indicates a transition from an anti-clockwise trap-dominated hysteresis at low temperature to a clockwise ionic polarization dominated hysteresis for temperatures above 260 K. Note that the temperature corresponding to the change in direction of hysteresis is similar to the temperature corresponding to the onset of ionic diffusion (~ 270 K ± 10 K) obtained in **Figure 1d**, indicating that ionic diffusion is observed only above 260 K.[47, 52]

Further evidence for the existence of ionic polarization at the SAND-semiconductor interface is obtained from the sweep rate variation of the transfer measurements for the SAND-based PFETs. In trap-limited transport, $\Delta V_{th}$ and the area between hysteresis curves decreases with slowing sweep rate, due to charge carrier trapping and detrapping.[48, 57-58] However, when the transport has a contribution from ionic polarization, an increase in $|\Delta V_{th}|$ is observed with a decrease in the $V_g$ sweep rate predominantly due to the relaxation mechanism of the ions at the dielectric-semiconductor interface.[42, 52] As is evident from **Figure 2**, at 300 K $|\Delta V_{th}|$ increases from near negligible values (below the step size used for the $V_g$ sweep) to up to 450 ± 150 mV when the sweep rate is decreased from 0.5 V/s to 0.05V/s. The observed variation in the hysteresis is significant when the devices are operated close to room temperature versus when they are measured at 100 K (**Figure S3,S4**), a classic signature of ionic transport.[42, 47]

Next, we examine the photo-excitation of the SAND dielectric for reversibly modulating the polarization and channel current. The SAND-semiconductor interface was illuminated with a UV source (λ ~ 365 nm, P ~ 14 mW/cm$^2$) and visible light sources (λ > 400 nm, P ~ 2 mW/cm$^2$) while carrying out transistor or capacitance measurements. A schematic illustration of the *trans*-<u>cis</u> photo-isomerization of the azo benzene[34] based PAE moiety is shown in **Figure 3a** and the corresponding optical absorption variation is shown in **Figure S5**. DFT simulations based on B3LYP hybrid functional and 6-31G(d,p) implemented in a Gaussian 09 suite estimate the dipole moment of a single polar PAE molecule changes from 14.46 D in the *trans*-state to the 8.65 D in the *cis*-state upon photoisomerization. Correspondingly, the dimension of the PAE molecule changes from 16.14 Å to 7.49 Å (details in **Figure S6**). Consequently, the thickness variation due to geometrical isomerization of the SAND structure under photo-isomerization would essentially compensate the change in the dipole moment, resulting in no expected variation in capacitance upon illumination if molecular dipoles were the only factor contributing to the polarization of the SAND structure.

We then measured the capacitance of the SAND structure using a metal-insulator-metal structure consisting of an ITO/SAND/Al trilayer. Typical capacitive response of the SAND layer upon illumination with UV and visible wavelengths is shown in **Figures 3b** and **S7**. We observe that the capacitance of the SAND layer can be reversibly increased 3-fold (0.11 μF/cm$^2$ to 0.34 μF/cm$^2$) and decreased (0.34 μF/cm$^2$ to 0.14 μF/cm$^2$) upon illumination with UV and visible wavelengths, respectively (**Figure 3c**) in apparent contradiction of the DFT calculations. This variation in the capacitance indicates that there are additional mechanisms responsible for the photo-response. To confirm the electrical integrity of the SAND superlattices, the displacement/leakage current was monitored upon continuous visible (λ > 400 nm) and UV (λ ~ 365 nm) illumination for 30 minutes (**Figure S7**), which indicates no evidence for breakdown of the dielectric structure. However, the impedance loss spectroscopy measured under UV and visible illumination indicates a significant increase in the loss corresponding to low frequency ionic relaxation, which is consistent with an increased ionic movement under UV illumination (**Figure S7b**). Therefore, the observed variation in capacitance can be attributed to the change in the ionic polarization due to the geometrical rearrangement of the PAE molecular dipoles under illumination. Scanning capacitance microscopy (SCM) measurement of the SAND layer performed upon illumination indicates an overall reversible modification of the surface polarization upon UV illumination (**Figure S8**). The SCM measurement only accounts for a 10% change in polarization. This can possibly be attributed to the fact that the total polarization originates from the combination of the bulk and surface polarization (details in Supplementary Information page **S4**).

In addition, the SAND-based PFETs show an increase in $I_{ds}$ upon illumination with visible light due to an increase in photo-generated charges (**Figure 4a**). Correspondingly we observe a change in threshold voltage from 1.2 V in dark to 1.4 V on visible illumination and 1.6 V under UV illumination. This slight variation in the $V_{th}$ can be attributed to the photo-induced charges and ionic polarization, respectively. The effect of dipolar disorder induced changes in the $V_{th}$ is not observed because of the presence of ODPA interlayer and smooth SAND-semiconductor interface.[11] Note that, with the choice of wavelength for UV illumination, it was ensured that the p-type polymer PBTOR has a minimal absorption at 365nm (**Figure S9**). Upon UV illumination, the SAND-based PFETs show an obvious increase in hysteresis at 300 K, which is not present when the same devices are UV illuminated at 100 K (**Figure 4b**), indicating the dominant effect of ionic polarization. In general, an increase in the hysteresis upon UV illumination in PFETs has been attributed to UV-induced trap formation in the transport layer[59-61]. Hence, in order to isolate potential artefacts which can arise due to UV illumination on the p-type polymer, a control measurement was performed on BCB dielectric layer based p-FETs (**Figure S10**). A much smaller enhancement (~15%) in the $I_{ds}$ is observed upon UV illumination, in accord with the low absorption coefficient for the particular wavelength of 365 nm (**Figure S9**). In contrast, upon visible light illumination, a much larger enhancement in the channel current of up to 200%, with a typical R ~ 12 A/W, is observed for the BCB-based p-FETs. Nevertheless, the devices exhibit excellent stability in the channel current even after continuous UV illumination for more than 30 minutes. Hence, it can be safely concluded that UV-based generation of traps or other potential damage to PBTOR can be neglected in the present case (**Figure S10, S11**). Moreover, unlike the SAND-based p-FETs, the hysteresis behavior is not altered upon UV or visible illumination for the BCB-based p-FETs. Hence, it appears likely that UV illumination isomerizes the PAE to the *cis*-isomer, which geometrically increases the diffusivity of the Br$^-$ ions and thus, the transfer curve hysteresis. The intrinsic photosensitivity of the SAND-based p-FETs under $V_g$ variation with visible and UV illumination is shown in **Figure 4c**. After compensating for the photoresponse due to photogenerated carriers generated by UV absorption, we estimate R values for a six-layer SAND (thickness ~ tens of nms) given by the expression: $(I_d^{UV+Isomerization} - I_d^{UV})/P_{ill}$ to be ~ 2 A/W , where $I_d^{UV}$ corresponds to the channel current obtained upon UV illumination for BCB-PBTOR devices and $I_d^{UV+Isomerization}$ corresponds to the channel current under UV illumination for SAND-PBTOR devices. This approach to obtain the effect of isomerisation on the photo-response is reasonable since similar visible photoresponse of ~ 10 – 12 A/W is observed on both SAND-PBTOR as well as BCB-PBTOR devices (**Figure S10**).

Upon UV illumination, the transfer curve gradually shifts and saturates after 40 minutes, presumably reaching an equilibrium state where photo-isomerization is no longer possible (**Figure 4d**). The observed variation in the transfer characteristics can then be related to the associated ionic polarization change due to the PAE photo-isomerization which in turn increases the Br$^-$ ion diffusivity. On the other hand, when the devices were illuminated with visible light, reverse isomerization increases and $I_{ds}$ decreases with an observed decrease in hysteresis. The decrease in the channel current occurs until the point in which the $I_{ds}$ is dominated by excess photogenerated charge carriers. Hence, complete reversibility to the dark current levels is not possible with Al gated SAND- FETs. Based on the time evolution of the channel current under UV and visible illumination (**Figure S12b**) it is possible to estimate the fraction of photoisomerization to be 54.5% (details in Supporting Information **S7**). Nevertheless, to ensure complete reversibility we fabricated ITO-gated SAND FETs which allow illuminating the device from dielectric side. **Figure 4e** depicts the transfer characteristics measured under UV, visible and dark conditions which exhibits essentially complete reversibility. This lack of reversibility in SAND-FET system due to UV and visible illumination is a limitation of this system which originates in the ionic relaxation mechanism of the SAND system (detailed discussion in Supporting Information page **S7**). When the pristine devices (SAND is thermodynamically stable in the *trans* configuration) are exposed to visible illumination that involves no isomerization, the reversibility of the transfer curves is instantaneous at 300 K (**Figure 4f**). On the basis of this property, SAND-based PFETs are well-suited for non-invasive light-based multi-state memory devices. However, the ability to realize non-volatile memory effects in these photo-isomeric based PFETs is limited by the thermodynamics of the photo-isomerization process and the associated Br$^-$ ion relaxation which exhibits typical Arrhenius behavior.[24] Nevertheless, as a proof-of-concept, **Figure S12** shows a series of transfer plots obtained with dark, visible, and UV illumination that correspond to different SAND-PFET logic states.

**Conclusions**

Here we utilize self-assembly chemistry to design photo-tunable nanodielectric films. These nanodielectric films exhibit reversible capacitance modulation from 0.1 µF/cm$^2$ to 0.3 µF/cm$^2$ upon photo isomerization. Photo-tunable nanodielectrics are then used as dielectric layers in PFETs demonstrating $\mu_{FET}$ > 2 cm$^2$V$^{-1}$s$^{-1}$ at low operating voltages of $V_d = V_g = 5$ V and a strong photoresponsivity of up to 2 A/W originating from the isomerization of the organic phase in the nanodielectric films. This strategy enables the isolation and differentiation of ionic, molecular, and atomic polarization contributions at the dielectric/semiconductor interface in the same FET device and also enables multifunctional bistable optoelectronic memory devices.

**Experimental Section**

*Materials: Conjugated Polymers PBTOR ($M_n$ = 5.2 kDa, PDI = 3.2) and N2200 ($M_w$ ≈ 180 kDA, PDI = 3.89) were procured from Polyera Corp., USA. The precursor for $HfO_x$ dielectric fabrication, $HfCl_4$, and n-octadecylphosphonic acid (ODPA) for surface modification were obtained from Sigma Aldrich Inc. ITO coated glass slides were obtained from Xinyan Technologies Ltd.*

*FET fabrication: Bottom-gate/top-contact devices were fabricated by coating patterned Al ($10^{-6}$ mbar, 1 A°/s, 30 nm thick) gate electrodes on pre-cleaned RCA glass substrates. For transparent PFETs, ITO was patterned and used as the gate electrode. This was followed by Hf-SAND growth (six layer iterative growth) as described in Reference 7a. The surface of the Hf-SAND layer was modified by self-assembled monolayers (SAMs) of ODPA. Phosphonic acid based SAMs were prepared by immersing the SAND coated substrates in a 2 mM ethanolic solution of ODPA overnight, rinsing in EtOH, and drying under a nitrogen stream. Polymer active layers (50 nm) PBTOR and N2200 were spin-coated from a 10 mg/mL solution in chlorobenzene at 1000 rpm for 1 min. The active layers were annealed in nitrogen atmosphere at $110^0$ C for 30 mins. This was followed by coating patterned aligned Au S-D electrodes ($10^{-6}$ mbar, 1 A°/s, 30 nm thick) to complete the device fabrication. Capacitance measurement was performed using a standard M-I-M structure where the metal layers were evaporated using the similar procedure as the FET fabrication.*

*Electrical Characterization: Standard PFET characterization was performed using a Keithley 4200 SCS, and the capacitance measurements were performed using a HP4294A unit. $\mu_{FET}(T)$ was estimated by varying the T using a He gas based set up from Cryogenics Technology Ltd. Photo-FET measurements were performed for samples inside the cryo chamber by illumination through the chamber watch glass with sources: UV LED (λ ~ 365 nm, P ~ 14 mW/cm$^2$) from Hamamtsu Co. and ultra bright LED (λ > 400 nm, P ~ 2 mW/cm$^2$) from Phillips. The reported power values are the power*

*measured through a standard Si detector when placed at the sample position in the cryostat. Since the isomerisation process is a dynamic process, unless otherwise specified, all the transfer curves shown are after near complete isomerisation (after which no further change in channel current is observed upon further illumination) which is generally observed in our case to be after 30-40 minutes.*

**Associated Content**

Supporting Information contains: SAND-N2200 FET characterization; temperature dependent transfer characteristics of n-type transistor; sweep rate variation at 100 K; DFT, impedance and spectroscopic characterization of photoisomerization; absorption spectra of PBTOR; transistor characteristics of BCB-PBTOR devices; optoelectronic memory in SAND-PFETs. This material is available free of charge via the Internet at http://pubs.acs.org.

**Author Information:**

Corresponding Authors: narayan@jncasr.ac.in, t-marks@northwestern.edu, m-hersam@northwestern.edu, a-facchetti@northwestern.edu.

**Notes:** The authors declare no competing financial interest.

**Acknowledgements**

SPS acknowledges Dr. Chidambar Kulkarni and Prof. E.W.Meijer, Eindhoven University of Technology for stimulating discussions of photochromism. SPS acknowledges funding from CSIR-HRDG. KSN acknowledges funding from DAE, India. This research was also supported by IUSSTF Funding, and by the National Science Foundation through the Northwestern University Materials Research Science and Engineering Center (NSF DMR-1720139).

**Figures:**

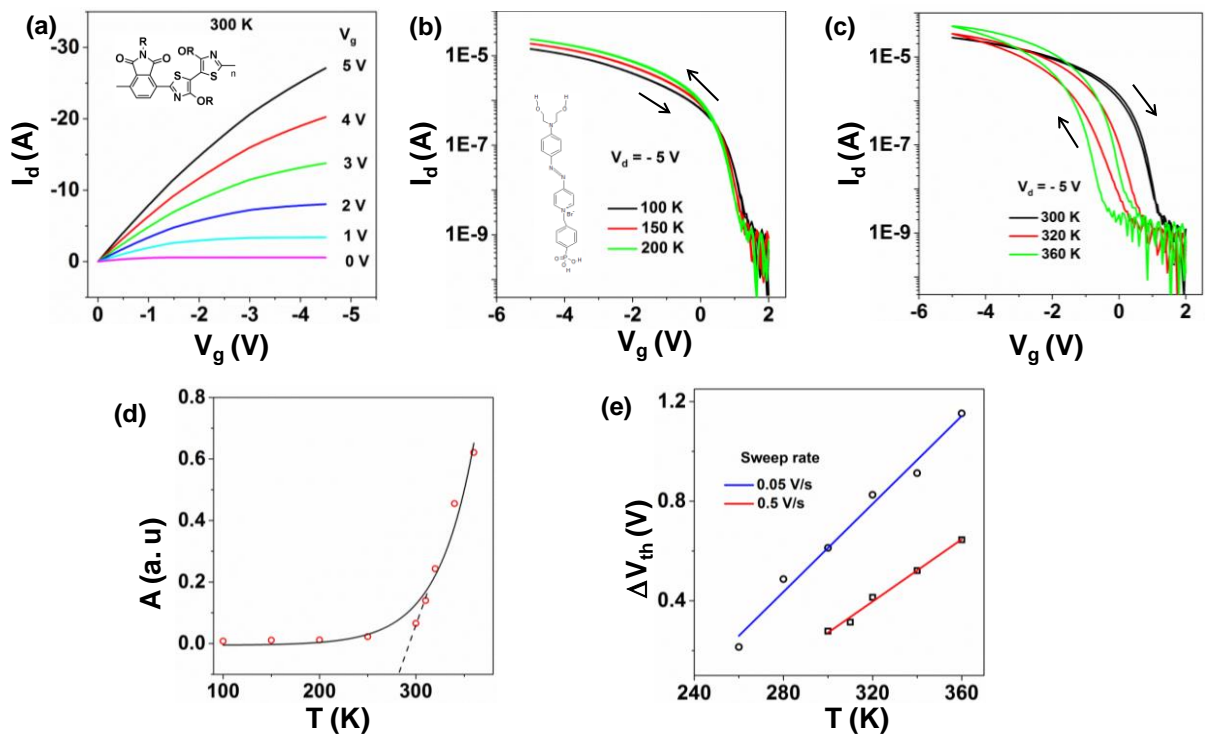

**Figure 1.** (a) Output characteristics of SAND based p-FETs (L = 60 μm, W = 1 mm) fabricated with the PBTOR semiconductor; inset shows the chemical structure of PBTOR. (b) Low temperature (inset shows the structure of PAE moiety) FET transfer characteristics; (c) high temperature transfer characteristics of same FET measured at a sweep rate of 0.5V/s; (d) variation of area under the transfer curves "A" for the same FET when measured with a sweep rate of 0.5 V/s; (e) variation of threshold voltage difference between the forward and reverse sweep with temperature for different sweep rates of $V_g$.

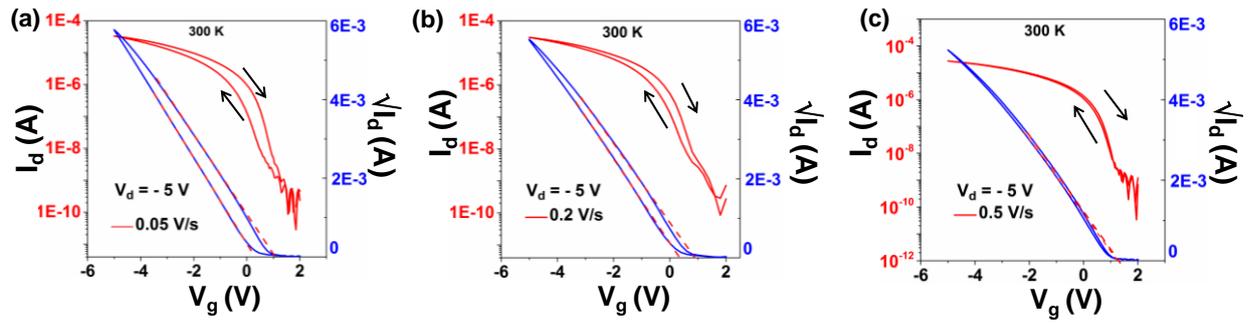

**Figure 2.** Effect of the variation of $V_g$ sweep rate on the hysteresis of the transfer characteristics of SAND-based p-FETs (a) 0.05 V/s; (b) 0.2 V/s; (c) 0.5 V/s. Also shown is the linear fit to $I_d^{0.5}$ with $V_g$ indicating that the $V_{th}$ shift corresponds to the forward and reverse sweep.

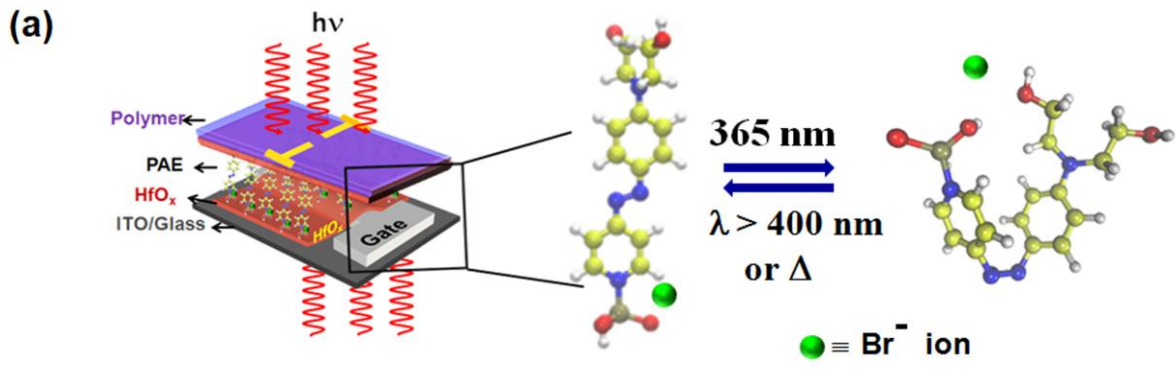
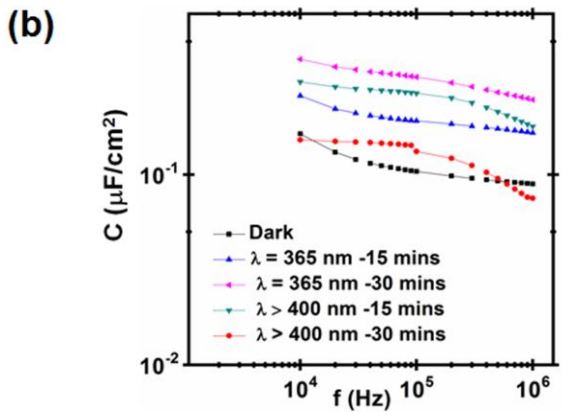
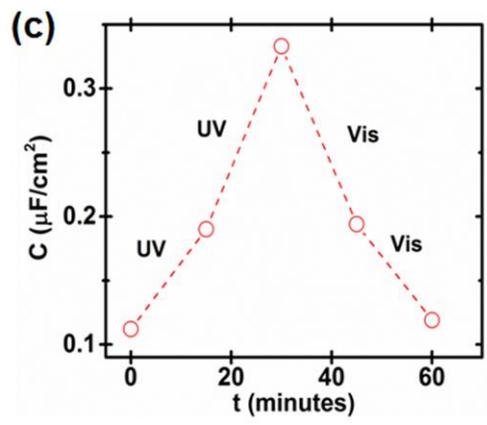

**Figure 3.** (a) Schematic of photo-transistor and the mechanism of the *trans-cis* photo-isomerization of the SAND layer. (b) Capacitance versus frequency behavior indicating the variation in capacitance upon illumination measured using a metal-insulator-metal structure. (c) Magnitude of the capacitance variation at 40 kHz under UV and visible illumination.

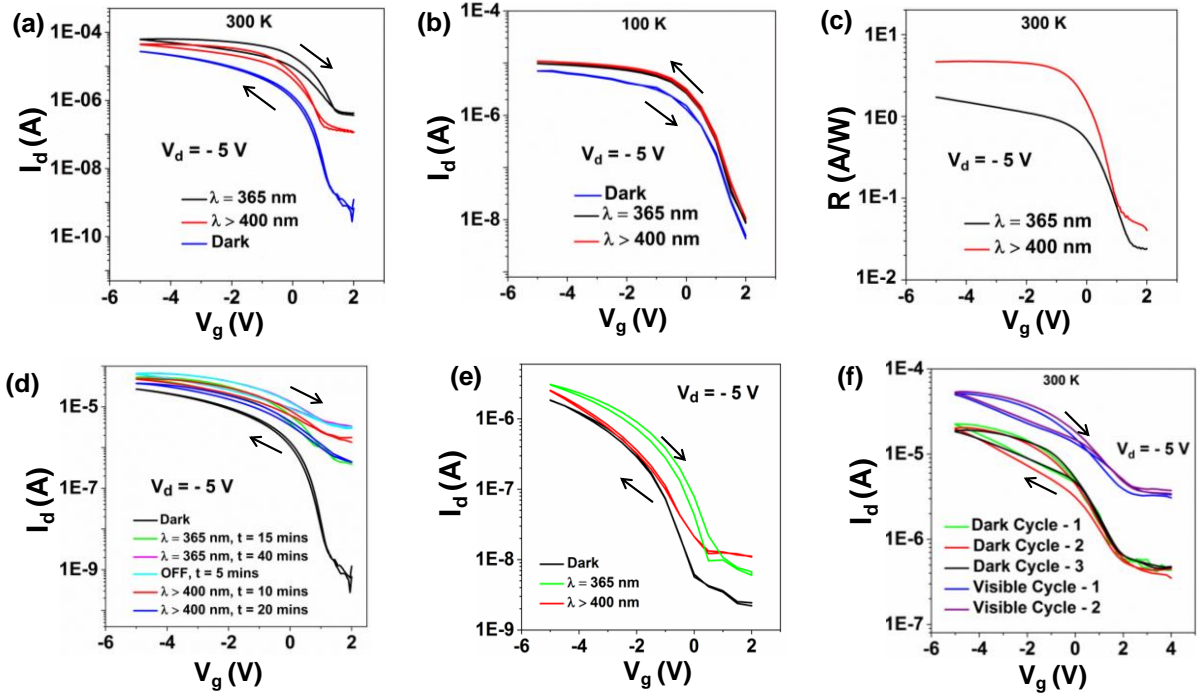

**Figure 4:** Variation in the hysteresis behaviour of transfer characteristics measured at 0.5 V/s upon visible (λ > 400 nm) and UV illumination (λ = 365 nm) at (a) 300 K and (b) 100 K on a different SAND-PBTOR device (L = 60 μm, W = 1 mm). (c) Responsivity as a function of $V_g$ upon visible (λ > 400 nm) and UV illumination (λ = 365 nm) of the same device. (d) Time evolution of the transfer characteristics upon UV illumination and consequent visible illumination indicating the effect of the isomerisation process. (e) Variation in the hysteresis behaviour of transfer characteristics measured at 0.5 V/s upon visible (λ > 400 nm) and UV illumination (λ = 365 nm) depicting the effect of forward and reverse isomerisation. (f) Reversibility of the FET response upon visible illumination "Visible Cycle" and switching off cycles "Dark Cycle".

**References**

(1) Sirringhaus, H. 25th Anniversary Article: Organic Field-Effect Transistors: The Path Beyond Amorphous Silicon. *Adv. Mater.* **2014,** *26* (9), 1319-1335, DOI: 10.1002/adma.201304346.
(2) Ha, Y.-G.; Everaerts, K.; Hersam, M. C.; Marks, T. J. Hybrid Gate Dielectric Materials for Unconventional Electronic Circuitry. *Acc. Chem. Res.* **2014,** *47* (4), 1019-1028, DOI: 10.1021/ar4002262.
(3) Kanimozhi, C.; Yaacobi-Gross, N.; Chou, K. W.; Amassian, A.; Anthopoulos, T. D.; Patil, S. Diketopyrrolopyrrole–Diketopyrrolopyrrole-Based Conjugated Copolymer for High-Mobility Organic Field-Effect Transistors. *J. Am. Chem. Soc.* **2012,** *134* (40), 16532-16535, DOI: 10.1021/ja308211n.
(4) Nielsen, C. B.; Turbiez, M.; McCulloch, I. Recent Advances in the Development of Semiconducting DPP-Containing Polymers for Transistor Applications. *Adv. Mater.* **2013,** *25* (13), 1859-1880, DOI: 10.1002/adma.201201795.
(5) Senanayak, S. P.; Ashar, A. Z.; Kanimozhi, C.; Patil, S.; Narayan, K. S. Room-temperature bandlike transport and Hall effect in a high-mobility ambipolar polymer. *Physical Review B* **2015,** *91* (11), 115302.
(6) Senanayak, S. P.; Narayan, K. S. Strategies for Fast-Switching in All-Polymer Field Effect Transistors. *Adv. Funct. Mater.* **2014,** *24* (22), 3324-3331, DOI: 10.1002/adfm.201303374.
(7) Herlogsson, L.; Noh, Y.-Y.; Zhao, N.; Crispin, X.; Sirringhaus, H.; Berggren, M. Downscaling of Organic Field-Effect Transistors with a Polyelectrolyte Gate Insulator. *Adv. Mater.* **2008,** *20* (24), 4708-4713, DOI: 10.1002/adma.200801756.
(8) Ikawa, M.; Yamada, T.; Matsui, H.; Minemawari, H.; Tsutsumi, J. y.; Horii, Y.; Chikamatsu, M.; Azumi, R.; Kumai, R.; Hasegawa, T. Simple push coating of polymer thin-film transistors. *Nature Communications* **2012,** *3*, 1176, DOI: 10.1038/ncomms2190.
(9) Yuan, Y.; Giri, G.; Ayzner, A. L.; Zoombelt, A. P.; Mannsfeld, S. C. B.; Chen, J.; Nordlund, D.; Toney, M. F.; Huang, J.; Bao, Z. Ultra-high mobility transparent organic thin film transistors grown by an off-centre spin-coating method. *Nature Communications* **2014,** *5*, 3005, DOI: 10.1038/ncomms4005.
(10) Giri, G.; Li, R.; Smilgies, D.-M.; Li, E. Q.; Diao, Y.; Lenn, K. M.; Chiu, M.; Lin, D. W.; Allen, R.; Reinspach, J.; Mannsfeld, S. C. B.; Thoroddsen, S. T.; Clancy, P.; Bao, Z.; Amassian, A. One-dimensional self-confinement promotes polymorph selection in large-area organic semiconductor thin films. *Nature Communications* **2014,** *5*, 3573, DOI: 10.1038/ncomms4573.
(11) Senanayak, S. P.; Sangwan, V. K.; McMorrow, J. J.; Everaerts, K.; Chen, Z.; Facchetti, A.; Hersam, M. C.; Marks, T. J.; Narayan, K. S. Self-Assembled Nanodielectrics for High-Speed, Low-Voltage Solution-Processed Polymer Logic Circuits. *Advanced Electronic Materials* **2015,** *1* (12), 1500226, DOI: 10.1002/aelm.201500226.
(12) Noh, Y.-Y.; Zhao, N.; Caironi, M.; Sirringhaus, H. Downscaling of self-aligned, all-printed polymer thin-film transistors. *Nat Nano* **2007,** *2* (12), 784-789, DOI: http://www.nature.com/nnano/journal/v2/n12/suppinfo/nnano.2007.365_S1.html.
(13) Higgins, S. G.; Muir, B. V. O.; Dell; apos; Erba, G.; Perinot, A.; Caironi, M.; Campbell, A. J. Self-aligned organic field-effect transistors on plastic with picofarad overlap capacitances and megahertz operating frequencies. *Applied Physics Letters* **2016,** *108* (2), 023302, DOI: 10.1063/1.4939045.
(14) DiBenedetto, S. A.; Facchetti, A.; Ratner, M. A.; Marks, T. J. Molecular Self-Assembled Monolayers and Multilayers for Organic and Unconventional Inorganic Thin-Film Transistor Applications. *Adv. Mater.* **2009,** *21* (14-15), 1407-1433, DOI: 10.1002/adma.200803267.
(15) Narayan, K. S.; Kumar, N. Light responsive polymer field-effect transistor. *Applied Physics Letters* **2001,** *79* (12), 1891-1893, DOI: 10.1063/1.1404131.
(16) Zaumseil, J.; Friend, R. H.; Sirringhaus, H. Spatial control of the recombination zone in an ambipolar light-emitting organic transistor. *Nat Mater* **2006,** *5* (1), 69-74, DOI: http://www.nature.com/nmat/journal/v5/n1/suppinfo/nmat1537_S1.html.
(17) Zhang, C.; Chen, P.; Hu, W. Organic Light-Emitting Transistors: Materials, Device Configurations, and Operations. *Small* **2016,** *12* (10), 1252-1294, DOI: 10.1002/smll.201502546.
(18) Lin, P.; Yan, F. Organic Thin-Film Transistors for Chemical and Biological Sensing. *Advanced Materials* **2012,** *24* (1), 34-51, DOI: 10.1002/adma.201103334.

(19) Kim, S.-J.; Lee, J.-S. Flexible Organic Transistor Memory Devices. *Nano Letters* **2010**, *10* (8), 2884-2890, DOI: 10.1021/nl1009662.
(20) Orgiu, E.; Crivillers, N.; Herder, M.; Grubert, L.; Pätzel, M.; Frisch, J.; Pavlica, E.; Duong, D. T.; Bratina, G.; Salleo, A.; Koch, N.; Hecht, S.; Samorì, P. Optically switchable transistor via energy-level phototuning in a bicomponent organic semiconductor. *Nat Chem* **2012**, *4* (8), 675-679, DOI: http://www.nature.com/nchem/journal/v4/n8/abs/nchem.1384.html#supplementary-information.
(21) Gemayel, M. E.; Börjesson, K.; Herder, M.; Duong, D. T.; Hutchison, J. A.; Ruzié, C.; Schweicher, G.; Salleo, A.; Geerts, Y.; Hecht, S.; Orgiu, E.; Samorì, P. Optically switchable transistors by simple incorporation of photochromic systems into small-molecule semiconducting matrices. *Nature Communications* **2015**, *6*, 6330, DOI: 10.1038/ncomms7330.
(22) Frolova, L. A.; Troshin, P. A.; Susarova, D. K.; Kulikov, A. V.; Sanina, N. A.; Aldoshin, S. M. Photoswitchable organic field-effect transistors and memory elements comprising an interfacial photochromic layer. *Chemical Communications* **2015**, *51* (28), 6130-6132, DOI: 10.1039/c5cc00711a.
(23) Tsuruoka, T.; Hayakawa, R.; Kobashi, K.; Higashiguchi, K.; Matsuda, K.; Wakayama, Y. Laser Patterning of Optically Reconfigurable Transistor Channels in a Photochromic Diarylethene Layer. *Nano Letters* **2016**, DOI: 10.1021/acs.nanolett.6b03162.
(24) Zhang, H.; Guo, X.; Hui, J.; Hu, S.; Xu, W.; Zhu, D. Interface Engineering of Semiconductor/Dielectric Heterojunctions toward Functional Organic Thin-Film Transistors. *Nano Letters* **2011**, *11* (11), 4939-4946, DOI: 10.1021/nl2028798.
(25) Körner, P. O.; Shallcross, R. C.; Maibach, E.; Köhnen, A.; Meerholz, K. Optical and electrical multilevel storage in organic memory passive matrix arrays. *Organic Electronics* **2014**, *15* (12), 3688-3693, DOI: https://doi.org/10.1016/j.orgel.2014.10.011.
(26) Crivillers, N.; Liscio, A.; Di Stasio, F.; Van Dyck, C.; Osella, S.; Cornil, D.; Mian, S.; Lazzerini, G. M.; Fenwick, O.; Orgiu, E.; Reinders, F.; Braun, S.; Fahlman, M.; Mayor, M.; Cornil, J.; Palermo, V.; Cacialli, F.; Samori, P. Photoinduced work function changes by isomerization of a densely packed azobenzene-based SAM on Au: a joint experimental and theoretical study. *Physical Chemistry Chemical Physics* **2011**, *13* (32), 14302-14310, DOI: 10.1039/c1cp20851a.
(27) Crivillers, N.; Orgiu, E.; Reinders, F.; Mayor, M.; Samorì, P. Optical Modulation of the Charge Injection in an Organic Field-Effect Transistor Based on Photochromic Self-Assembled-Monolayer-Functionalized Electrodes. *Advanced Materials* **2011**, *23* (12), 1447-1452, DOI: 10.1002/adma.201003736.
(28) Shallcross, R. C.; Körner, P. O.; Maibach, E.; Köhnen, A.; Meerholz, K. A Photochromic Diode With a Continuum of Intermediate States: Towards High Density Multilevel Storage. *Advanced Materials* **2013**, *25* (34), 4807-4813, DOI: doi:10.1002/adma.201301286.
(29) Meng, X.; Gui, B.; Yuan, D.; Zeller, M.; Wang, C. Mechanized azobenzene-functionalized zirconium metal-organic framework for on-command cargo release. *Science Advances* **2016**, *2* (8), DOI: 10.1126/sciadv.1600480.
(30) Orgiu, E.; Samorì, P. 25th Anniversary Article: Organic Electronics Marries Photochromism: Generation of Multifunctional Interfaces, Materials, and Devices. *Adv. Mater.* **2014**, *26* (12), 1827-1845, DOI: 10.1002/adma.201304695.
(31) Pärs, M.; Hofmann, C. C.; Willinger, K.; Bauer, P.; Thelakkat, M.; Köhler, J. An Organic Optical Transistor Operated under Ambient Conditions. *Angewandte Chemie International Edition* **2011**, *50* (48), 11405-11408, DOI: doi:10.1002/anie.201104193.
(32) Arnold, H. N.; Cress, C. D.; McMorrow, J. J.; Schmucker, S. W.; Sangwan, V. K.; Jaber-Ansari, L.; Kumar, R.; Puntambekar, K. P.; Luck, K. A.; Marks, T. J.; Hersam, M. C. Tunable Radiation Response in Hybrid Organic–Inorganic Gate Dielectrics for Low-Voltage Graphene Electronics. *ACS Applied Materials & Interfaces* **2016**, *8* (8), 5058-5064, DOI: 10.1021/acsami.5b12259.
(33) Kang, H.; Evmenenko, G.; Dutta, P.; Clays, K.; Song, K.; Marks, T. J. X-Shaped Electro-optic Chromophore with Remarkably Blue-Shifted Optical Absorption. Synthesis, Characterization, Linear/Nonlinear Optical Properties, Self-Assembly, and Thin Film Microstructural Characteristics. *Journal of the American Chemical Society* **2006**, *128* (18), 6194-6205, DOI: 10.1021/ja060185v.
(34) Keinan, S.; Ratner, M. A.; Marks, T. J. Self-Assembled Electrooptic Superlattices. A Theoretical Study of Multilayer Formation and Response Using Donor−Acceptor, Hydrogen-Bond Building Blocks. *Chemistry of Materials* **2004**, *16* (10), 1848-1854, DOI: 10.1021/cm0300720.

(35) Lin, Q.; Pearson, R. A.; Hedrick, J. C. *Polymers for Microelectronics and Nanoelectronics*, American Chemical Society: 2004; Vol. 874, p 356.
(36) Senanayak, S. P.; Guha, S.; Narayan, K. S. Polarization fluctuation dominated electrical transport processes of polymer-based ferroelectric field effect transistors. *Phys. Rev. B* **2012,** *85* (11), 115311.
(37) Veres, J.; Ogier, S. D.; Leeming, S. W.; Cupertino, D. C.; Mohialdin Khaffaf, S. Low-k Insulators as the Choice of Dielectrics in Organic Field-Effect Transistors. *Advanced Functional Materials* **2003,** *13* (3), 199-204, DOI: 10.1002/adfm.200390030.
(38) Hulea, I. N.; Fratini, S.; Xie, H.; Mulder, C. L.; Iossad, N. N.; Rastelli, G.; Ciuchi, S.; Morpurgo, A. F. Tunable Frohlich polarons in organic single-crystal transistors. *Nat Mater* **2006,** *5* (12), 982-986, DOI: http://www.nature.com/nmat/journal/v5/n12/suppinfo/nmat1774_S1.html.
(39) Luzio, A.; Ferré, F. G.; Fonzo, F. D.; Caironi, M. Hybrid Nanodielectrics for Low-Voltage Organic Electronics. *Advanced Functional Materials* **2014,** *24* (12), 1790-1798, DOI: 10.1002/adfm.201302428.
(40) Li, L.; Lu, N.; Liu, M. Effect of dipole layer on the density-of-states and charge transport in organic thin film transistors. *Applied Physics Letters* **2013,** *103* (25), 253303, DOI: 10.1063/1.4852137.
(41) Li, L.; Lu, N.; Liu, M.; Bässler, H. General Einstein relation model in disordered organic semiconductors under quasiequilibrium. *Physical Review B* **2014,** *90* (21), 214107, DOI: 10.1103/PhysRevB.90.214107.
(42) Kim, S. H.; Hong, K.; Xie, W.; Lee, K. H.; Zhang, S.; Lodge, T. P.; Frisbie, C. D. Electrolyte-Gated Transistors for Organic and Printed Electronics. *Advanced Materials* **2013,** *25* (13), 1822-1846, DOI: 10.1002/adma.201202790.
(43) Inganas, O. Hybrid electronics and electrochemistry with conjugated polymers. *Chemical Society Reviews* **2010,** *39* (7), 2633-2642, DOI: 10.1039/b918146f.
(44) Everaerts, K.; Emery, J. D.; Jariwala, D.; Karmel, H. J.; Sangwan, V. K.; Prabhumirashi, P. L.; Geier, M. L.; McMorrow, J. J.; Bedzyk, M. J.; Facchetti, A.; Hersam, M. C.; Marks, T. J. Ambient-Processable High Capacitance Hafnia-Organic Self-Assembled Nanodielectrics. *J. Am. Chem. Soc.* **2013,** *135* (24), 8926-8939, DOI: 10.1021/ja4019429.
(45) DiBenedetto, S. A.; Facchetti, A.; Ratner, M. A.; Marks, T. J. Charge Conduction and Breakdown Mechanisms in Self-Assembled Nanodielectrics. *J. Am. Chem. Soc.* **2009,** *131* (20), 7158-7168, DOI: 10.1021/ja9013166.
(46) Orgiu, E.; Locci, S.; Fraboni, B.; Scavetta, E.; Lugli, P.; Bonfiglio, A. Analysis of the hysteresis in organic thin-film transistors with polymeric gate dielectric. *Organic Electronics* **2011,** *12* (3), 477-485, DOI: http://dx.doi.org/10.1016/j.orgel.2010.12.014.
(47) Panzer, M. J.; Frisbie, C. D. High Carrier Density and Metallic Conductivity in Poly(3-hexylthiophene) Achieved by Electrostatic Charge Injection. *Advanced Functional Materials* **2006,** *16* (8), 1051-1056, DOI: 10.1002/adfm.200600111.
(48) Egginger, M.; Bauer, S.; Schwödiauer, R.; Neugebauer, H.; Sariciftci, N. S. Current versus gate voltage hysteresis in organic field effect transistors. *Monatshefte für Chemie - Chemical Monthly* **2009,** *140* (7), 735-750, DOI: 10.1007/s00706-009-0149-z.
(49) Ucurum, C.; Goebel, H.; Yildirim, F. A.; Bauhofer, W.; Krautschneider, W. Hole trap related hysteresis in pentacene field-effect transistors. *Journal of Applied Physics* **2008,** *104* (8), 084501, DOI: doi:http://dx.doi.org/10.1063/1.2999643.
(50) Yoon, M.-H.; Kim, C.; Facchetti, A.; Marks, T. J. Gate Dielectric Chemical Structure−Organic Field-Effect Transistor Performance Correlations for Electron, Hole, and Ambipolar Organic Semiconductors. *Journal of the American Chemical Society* **2006,** *128* (39), 12851-12869, DOI: 10.1021/ja063290d.
(51) Tsai, T.-D.; Chang, J.-W.; Wen, T.-C.; Guo, T.-F. Manipulating the Hysteresis in Poly(vinyl alcohol)-Dielectric Organic Field-Effect Transistors Toward Memory Elements. *Advanced Functional Materials* **2013,** *23* (34), 4206-4214, DOI: 10.1002/adfm.201203694.
(52) Panzer, M. J.; Frisbie, C. D. Polymer Electrolyte-Gated Organic Field-Effect Transistors: Low-Voltage, High-Current Switches for Organic Electronics and Testbeds for Probing Electrical Transport at High Charge Carrier Density. *Journal of the American Chemical Society* **2007,** *129* (20), 6599-6607, DOI: 10.1021/ja0708767.
(53) Queffélec, C.; Petit, M.; Janvier, P.; Knight, D. A.; Bujoli, B. Surface Modification Using Phosphonic Acids and Esters. *Chemical Reviews* **2012,** *112* (7), 3777-3807, DOI: 10.1021/cr2004212.

(54) Panzer, M. J.; Frisbie, C. D. Exploiting Ionic Coupling in Electronic Devices: Electrolyte-Gated Organic Field-Effect Transistors. *Advanced Materials* **2008,** *20* (16), 3177-3180, DOI: 10.1002/adma.200800617.
(55) Paramonov, P. B.; Paniagua, S. A.; Hotchkiss, P. J.; Jones, S. C.; Armstrong, N. R.; Marder, S. R.; Brédas, J.-L. Theoretical Characterization of the Indium Tin Oxide Surface and of Its Binding Sites for Adsorption of Phosphonic Acid Monolayers. *Chemistry of Materials* **2008,** *20* (16), 5131-5133, DOI: 10.1021/cm8014622.
(56) Wang, S.; Ha, M.; Manno, M.; Daniel Frisbie, C.; Leighton, C. Hopping transport and the Hall effect near the insulator–metal transition in electrochemically gated poly(3-hexylthiophene) transistors. *Nature Communications* **2012,** *3*, 1210, DOI: 10.1038/ncomms2213.
(57) Se Hyun, K.; Won Min, Y.; Oh-Kwan, K.; Kipyo, H.; Chanwoo, Y.; Woon-Seop, C.; Chan Eon, P. Hysteresis behaviour of low-voltage organic field-effect transistors employing high dielectric constant polymer gate dielectrics. *Journal of Physics D: Applied Physics* **2010,** *43* (46), 465102.
(58) Kim, S. H.; Nam, S.; Jang, J.; Hong, K.; Yang, C.; Chung, D. S.; Park, C. E.; Choi, W.-S. Effect of the hydrophobicity and thickness of polymer gate dielectrics on the hysteresis behavior of pentacene-based field-effect transistors. *Journal of Applied Physics* **2009,** *105* (10), 104509, DOI: doi:http://dx.doi.org/10.1063/1.3131664.
(59) Lim, K. C.; Huang, W. F. Positron annihilation detection of ultra-violet light induced damage in conjugated polymers. *Solid State Communications* **1993,** *87* (9), 771-774, DOI: http://dx.doi.org/10.1016/0038-1098(93)90411-F.
(60) Krellner, C.; Haas, S.; Goldmann, C.; Pernstich, K. P.; Gundlach, D. J.; Batlogg, B. Density of bulk trap states in organic semiconductor crystals: Discrete levels induced by oxygen in rubrene. *Physical Review B* **2007,** *75* (24), 245115.
(61) Salleo, A. Electronic Traps in Organic Semiconductors. In *Organic Electronics*; Wiley-VCH Verlag GmbH & Co. KGaA: 2013; pp 341-380.